\documentclass[aps,twocolumn,showpacs,floatfix]{revtex4}
\usepackage{amsmath}
\usepackage{amsfonts}
\usepackage{amssymb}
\usepackage{epsfig}
\usepackage{graphicx}

\begin{document}
\title{Normal mode splitting in a coupled system of nanomechanical oscillator and parametric amplifier cavity}
\author{Sumei Huang and G. S. Agarwal}
\affiliation{Department of Physics, Oklahoma State University,
Stillwater, Oklahoma 74078, USA}
\date{\today}

\begin{abstract}
We study how an optical parametric amplifier inside the cavity can
affect the normal mode splitting behavior of the coupled movable
mirror and the cavity field. We work in the resolved sideband
regime. The spectra exhibit a double-peak structure as the
parametric gain is increased. Moreover, for a fixed parametric gain,
the double-peak structure of the spectrum is more pronounced with
increasing the input laser power. We give results for mode
splitting. The widths of the split lines are sensitive to parametric
gain.

\end{abstract}
 \maketitle

\renewcommand{\thesection}{\Roman{section}}
\setcounter{section}{0}
\section{Introduction}
\renewcommand{\baselinestretch}{1}\small\normalsize
    Recently there has been a major effort in applying many of the well tested
    ideas from quantum optics such as squeezing, quantum entanglement
    to optomechanical systems which are  macroscopic systems.
    Thus observation of entanglement~\cite{Knight,Marshall,DVitali,Paternostro,Bhattacharya1,Plenio},
    squeezing~\cite{Fabre,Mancini} etc in optomechanical systems
    would enable one to study quantum behavior at macroscopic scale.
    This of course requires cooling such systems to their ground state and significant
     advances have been made in cooling the mechanical mirror to far below the temperature of the
environment~\cite{Kleckner,Poggio,Gigan,Arcizet,Bhattacharya2,Bhattacharya3,Bhattacharya4}.
Further it has been pointed out that using optical back action one
can possibly achieve the ground state cooling in the resolved
sideband regime where the frequency of the mechanical mirror is much
larger than the cavity decay rate, that is
$\omega_{m}\gg\kappa$~\cite{Rae,Marquardt1,Schliesser}.

Another key idea from quantum optics is the vacuum Rabi
splitting~\cite{Eberly,Agarwal} which is due to strong interaction
between the atoms and the cavity mode. The experimentalists have
worked hard over the years to produce stronger and stronger
couplings to produce larger and larger
splittings~\cite{Raizen,Kimble,Rempe}. Application of these ideas to
macroscopic systems is challenging as well. In a recent paper
Kippenberg {\it et al.}~\cite{Kippenberg} proposed the possibility
of normal mode splitting in the resolved sideband regime using
optomechanical oscillators. In this paper, we propose placing a type
I optical parametric amplifier inside the cavity to increase the
coupling between the movable mirror and the cavity field, and this
should make the observation of the normal mode splitting of the
movable mirror and the output field more accessible.

The paper is structured as follows. In Sec. II we present the model,
derive the quantum Langevin equations, and give the steady-state
mean values. In Sec. III we present solution to the linearized
Langevin equations and give the spectrum of the movable mirror. In
Sec. IV we analyse and estimate the amount of the normal mode
splitting of the spectra. In Sec. V we calculate the spectra of the
output field. In Sec. VI we discuss the mode splitting of the
spectra of the movable mirror and the output field.
\section{Model}
\begin{figure}[htp]
 \scalebox{0.85}{\includegraphics{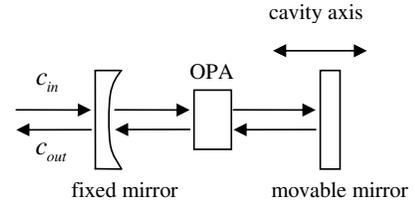}}
 \caption{\label{Fig1} Sketch of the studied system.
The cavity contains a nonlinear crystal which is pumped by a laser
(not shown) to produce parametric amplification and to change photon
statistics in the cavity.}
 \end{figure}

The system under consideration, sketched in Fig.~\ref{Fig1}, is an
optical parametric amplifier (OPA) placed within a Fabry-Perot
cavity formed by one fixed partially transmitting mirror and one
movable perfectly reflecting mirror in equilibrium with its
environment at a low temperature. The movable mirror is treated as a
quantum mechanical harmonic oscillator with effective mass $m$,
frequency $\omega_{m}$, and energy decay rate $\gamma_{m}$. An
external laser enters the cavity through the fixed mirror, then the
photons in the cavity will exert a radiation pressure force on the
movable mirror due to momentum transfer. This force is proportional
to the instantaneous photon number in the cavity.

In the adiabatic limit, the frequency $\omega_{m}$ of the movable
mirror is much smaller than the free spectral range of the cavity
$\frac{c}{2L}$ ($c$ is the speed of light in vacuum and $L$ is the
cavity length), the scattering of photons to other cavity modes can
be ignored, thus only one cavity mode $\omega_{c}$ is
considered~\cite{Lawf,Laws}. The Hamiltonian for the system in a
frame rotating at the laser frequency $\omega_{L}$ can be written as
\begin{eqnarray}\label{1}
H&=&\hbar(\omega_{c}-\omega_{L})n_{c}-\hbar\omega_{m}\chi n_{c}
Q+\frac{\hbar\omega_{m}}{4}(Q^2+P^2)\nonumber\\&
&+i\hbar\varepsilon(c^{\dag}-c)+i\hbar G (e^{i\theta}c^{\dag
2}-e^{-i\theta}c^{2}).
\end{eqnarray}

\noindent Here $Q$ and $P$ are the dimensionless position and
momentum operators for the movable mirror, defined by
$Q=\sqrt{\frac{2m\omega_{m}}{\hbar}}q$ and
$P=\sqrt{\frac{2}{m\hbar\omega_{m}}}p$ with $[Q,P]=2i$. In Eq.
(\ref{1}), the first term is the energy of the cavity field,
$n_{c}=c^{\dag}c$ is the number of the photons inside the cavity,
$c$ and $c^{\dag}$ are the annihilation and creation operators for
the cavity field satisfying the commutation relation $[c,c^\dag]=1$.
The second term comes from the coupling of the movable mirror to the
cavity field via radiation pressure, the dimensionless parameter
$\chi=\frac{1}{\omega_{m}}\frac{\omega_{c}}{L}\sqrt{\frac{\hbar}{2m\omega_{m}}}$
is the optomechanical coupling constant between the cavity and the
movable mirror. The third term corresponds the energy of the movable
mirror. The fourth term describes the coupling between the input
laser field and the cavity field, $\varepsilon$ is related to the
input laser power $\wp$ by $\varepsilon=\sqrt{\frac{2\kappa
\wp}{\hbar\omega_{L}}}$, where $\kappa$ is the cavity decay rate.
The last term is the coupling between the OPA and the cavity field,
\emph{G} is the nonlinear gain of the OPA, and $ \theta $ is the
phase of the field driving the OPA. The parameter \emph{G} is
proportional to the pump driving the OPA.

Using the Heisenberg equations of motion and adding the
corresponding damping and noise terms, we obtain the quantum
Langevin equations as follows,
\begin{equation}\label{2}
\begin{array}{lcl}
\dot{Q}=\omega_{m}P,\vspace*{.1in}\\
\dot{P}=2\omega_{m}\chi n_{c}-\omega_{m}Q-\gamma_{m}P+\xi,\vspace*{.1in}\\
\dot{c}=-i(\omega_{c}-\omega_{L}-\omega_{m}\chi
Q)c+\varepsilon+2Ge^{i\theta}c^{\dag}-\kappa c+\sqrt{2
\kappa}c_{in},\vspace*{.1in}\\
\dot{c}^{\dag}=i(\omega_{c}-\omega_{L}-\omega_{m}\chi
Q)c^{\dag}+\varepsilon+2Ge^{-i\theta}c-\kappa
c^{\dag}+\sqrt{2\kappa}c_{in}^{\dag}.
\end{array}
\end{equation}
Here we have introduced the input vacuum noise operator $c_{in}$
with zero mean value, which obeys the correlation function in the
time domain~\cite{Gardiner}
\begin{equation}\label{3}
\begin{array}{lcl}
\langle\delta c_{in}(t)\delta
c_{in}^{\dag}(t^{\prime})\rangle=\delta(t-t^{\prime}),\vspace*{.1in}\\
\langle\delta c_{in}(t)\delta
c_{in}(t^{\prime})\rangle=\langle\delta c_{in}^{\dag}(t)\delta
c_{in}(t^{\prime})\rangle=0.
\end{array}
\end{equation}
The force $\xi$ is the Brownian noise operator resulting from the
coupling of the movable mirror to the thermal bath, whose mean value
is zero, and it has the following correlation function at
temperature $T$~\cite{Giovannetti}:
\begin{equation}\label{4}
\langle \xi(t)\xi(t^{'})\rangle=\frac{1}{\pi}\frac{
\gamma_{m}}{\omega_{m}}\int\omega e^{-i\omega
(t-t^{'})}\left[1+\coth(\frac{\hbar \omega}{2k_{B}T})\right]d\omega,
\end{equation}
where $k_B$ is the Boltzmann constant and $T$ is the thermal bath
temperature. Following standard methods from quantum optics
~\cite{Walls}, we derive the steady-state solution to Eq. (\ref{2})
by setting all the time derivatives in Eq. (\ref{2}) to zero. They
are
\begin{equation}\label{5}
P_{s}=0,\hspace{.02in}Q_{s}=2\chi|c_{s}|^{2},\hspace{.02in}c_{s}=\frac{\kappa-i\Delta+2Ge^{i\theta}}{\kappa^{2}+\Delta^{2}-4G^{2}}\varepsilon,
\end{equation}
where
\begin{equation}\label{6}
\Delta=\omega_{c}-\omega_{L}-\omega_{m}\chi Q_{s}
\end{equation}
is the effective cavity detuning, depending on $Q_{s}$. The $Q_{s}$
denotes the new equilibrium position of the movable mirror relative
to that without the driving field. Further $c_{s}$ represents the
steady-state amplitude of the cavity field. From Eq. (\ref{5}) and
Eq. (\ref{6}), we can see $\Delta$ satisfies a fifth order equation,
it can at most have five real solutions. Therefore, the movable
mirror displays an optical multistable
behavior~\cite{Dorsel,Meystre,Marquardt}, which is a nonlinear
effect induced by the radiation-pressure coupling of the movable
mirror to the cavity field.

\section{Radiation pressure and quantum fluctuations}
In order to investigate the normal mode splitting of the movable
mirror and the output field, we need to calculate the fluctuations
of the system. Since the problem is nonlinear, we assume that the
nonlinearity is weak. Thus we can focus on the dynamics of small
fluctuations around the steady state of the system. Each operator of
the system can be written as the sum of its steady-state mean value
and a small fluctuation with zero mean value,
\begin{equation}\label{7}
Q=Q_{s}+\delta Q,\hspace*{.1in}P=P_{s}+\delta
P,\hspace*{.1in}c=c_{s}+\delta c.
\end{equation}
Inserting Eq. (\ref{7}) into Eq. (\ref{2}), then assuming
$|c_{s}|\gg1$, the linearized quantum Langevin equations for the
fluctuation operators take the form
\begin{equation}\label{8}
\begin{array}{lcl}
\delta\dot{Q}=\omega_{m}\delta P,\vspace*{.1in}\\
\delta\dot{P}=2\omega_{m}\chi (c^\ast_{s}\delta c+c_{s}\delta
c^{\dag})-\omega_{m}\delta Q-\gamma_{m}\delta P+\xi,\vspace{.1in}\\
\delta\dot{c}=-(\kappa+i\Delta)\delta c+i\omega_{m}\chi c_{s}\delta
Q+2Ge^{i\theta}\delta c^{\dag}+\sqrt{2 \kappa}\delta c_{in},\vspace*{.1in}\\
\delta\dot{c}^{\dag}=-(\kappa-i\Delta)\delta
c^{\dag}-i\omega_{m}\chi c^{*}_{s}\delta Q+2Ge^{-i\theta}\delta
c+\sqrt{2 \kappa}\delta c_{in}^{\dag}.
\end{array}
\end{equation}
Introducing the cavity field quadratures $\delta x=\delta c+\delta
c^{\dag}$ and $\delta y=i(\delta c^{\dag}-\delta c)$, and the input
noise quadratures $\delta x_{in}=\delta c_{in}+\delta c_{in}^{\dag}$
and $\delta y_{in}=i(\delta c_{in}^{\dag}-\delta c_{in})$, Eq.
(\ref{8}) can be rewritten in the matrix form
\begin{equation}\label{9}
\dot{f}(t)=Af(t)+\eta(t),
\end{equation}
in which $f(t)$ is the column vector of the fluctuations, $\eta(t)$
is the column vector of the noise sources. Their transposes are
\begin{equation}\label{10}
\begin{array}{lcl}
f(t)^{T}=(\delta Q,\delta P,\delta x,\delta y),\vspace*{.1in}\\
\eta(t)^{T}=(0,\xi,\sqrt{2\kappa}\delta x_{in},\sqrt{2\kappa}\delta y_{in});
\end{array}
\end{equation}
and the matrix $A$ is given by
\begin{equation}\label{11}
A=\left(
  \begin{array}{cccc}
    0 & \omega_{m} & 0 & 0 \vspace{.08in}\\
    -\omega_{m} & -\gamma_{m} & \omega_{m}\chi(c_{s}+c_{s}^{*}) & -i\omega_{m}\chi(c_{s}-c_{s}^{*}) \vspace{.08in}\\
    i\omega_{m}\chi (c_{s}-c_{s}^{*})  & 0 & 2G\cos\theta-\kappa & 2G\sin\theta+\Delta \vspace{.08in}\\
     \omega_{m}\chi (c_{s}+c_{s}^{*})  & 0 & 2G\sin\theta-\Delta & -(2G\cos\theta+\kappa) \vspace{.08in}\\
  \end{array}
\right).
\end{equation}
The system is stable only if all the eigenvalues of the matrix $A$
have negative real parts. The stability conditions for the system
can be derived by applying the Routh-Hurwitz
criterion~\cite{Hurwitz,DeJesus}. This gives
\begin{equation}\label{12}
\begin{array}{lcl}
2\kappa(\kappa^2-4G^{2}+\Delta^2+2\kappa\gamma_{m})+\gamma_{m}(2\kappa\gamma_{m}+\omega_{m}^2)>0,\vspace{.1in}\\
2\omega_{m}^{3}\chi^2(2\kappa+\gamma_{m})^2[|c_s|^2\Delta+iG(c_{s}^2e^{-i\theta}-c_{s}^{*2}e^{i\theta})]\vspace{.1in}\\\hspace{.25in}+\kappa\gamma_m
\{(\kappa^2-4G^2+\Delta^2)^2+(2\kappa\gamma_{m}+\gamma_{m}^2)\vspace{.1in}\\\hspace{.25in}\times(\kappa^2-4G^2+\Delta^2)
+\omega_{m}^2[2(\kappa^2+4G^2-\Delta^2)\vspace{.1in}\\\hspace{.25in}+\omega_{m}^2+2\kappa\gamma_{m}]\}>0,\vspace{.1in}\\
\kappa^{2}-4G^{2}+\Delta^2-4\omega_{m}\chi^2
[|c_s|^2\Delta+iG(c_{s}^2e^{-i\theta}-c_{s}^{*2}e^{i\theta})]>0.
\end{array}
\end{equation}
All the external parameters must be chosen to satisfy the stability
conditions (\ref{12}).

Taking Fourier transform of Eq. (\ref{8}) by using
$f(t)=\frac{1}{2\pi}\int_{-\infty}^{+\infty}f(\omega)e^{-i\omega t}
d\omega$ and
$f^{\dag}(t)=\frac{1}{2\pi}\int_{-\infty}^{+\infty}f^{\dag}(-\omega)e^{-i\omega
t} d\omega$, where $f^{\dag}(-\omega)=[f(-\omega)]^{\dag}$, then
solving it, we obtain the position fluctuations of the movable
mirror
\begin{equation}\label{13}
\begin{array}{lcl}
\delta
Q(\omega)=-\frac{\omega_{m}}{d(\omega)}[2\sqrt{2\kappa}\omega_m\chi\{[(\kappa-i(\Delta+\omega))c_{s}^{*}\vspace{.1in}\\\hspace{.5in}+2Ge^{-i\theta}c_{s}]\delta
c_{in}(\omega)
+[(\kappa+i(\Delta-\omega))c_{s}\vspace{.1in}\\\hspace{.5in}+2Ge^{i\theta}c_{s}^{*}]\delta c_{in}^{\dag}(-\omega)\}\vspace{.1in}\\
\hspace{.5in}+[(\kappa-i\omega)^2+\Delta^2-4G^2]\xi(\omega)],
\end{array}
\end{equation}
where
\begin{equation}\label{14}
\begin{array}{lcl}
d(\omega)=4\omega_{m}^3\chi^2[\Delta
|c_{s}|^2+iG(c_{s}^2e^{-i\theta}-c_{s}^{*2}e^{i\theta})]\vspace{.1in}\\\hspace{.5in}+(\omega^2-\omega_m^2+i\gamma_{m}\omega
)[(\kappa-i\omega)^2+\Delta^2-4G^2].
\end{array}
\end{equation}
In Eq. (\ref{13}), the first term proportional to $\chi$ originates
from radiation pressure, while the second term involving
$\xi(\omega)$ is from the thermal noise. So the position
fluctuations of the movable mirror are now determined by radiation
pressure and the thermal noise. In the case of no coupling with the
cavity field, the movable mirror will make Brownian motion, $\delta
Q(\omega)=\omega_{m}\xi(\omega)/(\omega_{m}^2-\omega^2-i\gamma_{m}\omega)$,
whose susceptibility has a Lorentzian shape centered at frequency
$\omega_{m}$ with width $\gamma_{m}$.

The spectrum of fluctuations in position of the movable mirror is
defined by
 \begin{equation}\label{15}
\frac{1}{2}(\langle\delta Q(\omega)\delta
Q(\Omega)\rangle+\langle\delta Q(\Omega)\delta
Q(\omega)\rangle)=2\pi S_{Q}(\omega)\delta(\omega+\Omega).
\end{equation}

To calculate the spectrum, we require the correlation functions of
the noise sources in the frequency domain,
\begin{equation}\label{16}
\begin{array}{lcl}
\langle\delta c_{in}(\omega)\delta
c_{in}^{\dag}(-\Omega)\rangle=2\pi\delta(\omega+\Omega),\vspace{.1in}\\
\langle\xi(\omega)\xi(\Omega)\rangle=4\pi
\frac{\gamma_{m}}{\omega_{m}}\omega\left[1+\coth(\frac{\hbar\omega}{2k_B
T})\right]\delta(\omega+\Omega).
\end{array}
\end{equation}
Substituting Eq. (\ref{13}) and Eq. (\ref{16}) into
 Eq. (\ref{15}), we obtain the spectrum of fluctuations in position of the movable
 mirror~\cite{Huang}
\begin{equation}\label{17}
\begin{array}{lcl}
S_{Q}(\omega)=\frac{\omega_{m}^2}{|d(\omega)|^2}\{8\omega_{m}^2\chi^2\kappa[(\kappa^2+\omega^2+\Delta^2+4G^2)|c_{s}|^2\vspace{.1in}\\
\hspace{.5in}+2Ge^{i\theta}c_{s}^{*2}(\kappa-i\Delta)+2Ge^{-i\theta}c_{s}^{2}(\kappa+i\Delta)]\vspace{.1in}\\
\hspace{.5in}+2\frac{\gamma_m}{\omega_{m}}\omega[(\Delta^2+\kappa^2-\omega^2-4G^2)^2+4\kappa^2\omega^2]\vspace{.1in}\\
\hspace{.5in}\times\coth(\frac{\hbar\omega}{2k_B T})\}.\\
\end{array}
\end{equation}
In Eq. (\ref{17}), the first term involving $\chi$ arises from
radiation pressure, while the second term originates from the
thermal noise. So the spectrum $S_{Q}(\omega)$ of the movable mirror
depends on radiation pressure and the thermal noise.

\section{Normal mode splitting and the eigenvalues of the matrix $A$}
The structure of all the spectra is determined by the eigenvalues of
$iA$ (Eq. (\ref{11})) or the complex zeroes of the function
$d(\omega)$ defined by Eq. (\ref{14}). Clearly we need the
eigenvalues of $iA$ as the solution of (Eq. (\ref{9})) in Fourier
domain is $f(\omega)=i(\omega-iA)^{-1}\eta(\omega)$. Let us analyse
the eigenvalues of Eq. (\ref{11}). Note that in the absence of the
coupling $\chi$=0, the eigenvalues of $iA$ are
\begin{equation}\label{18}
\begin{array}{lcl}
\pm\sqrt{\omega_{m}^2-\frac{\gamma_{m}^2}{4}}-\frac{i\gamma_{m}}{2};\pm\sqrt{\Delta^2-4G^2}-i\kappa.
\end{array}
\end{equation}
Thus the positive frequencies of the normal modes are given by
$\sqrt{\Delta^2-4G^2}$, $\sqrt{\omega_{m}^2-\frac{\gamma_{m}^2}{4}}$
$(\Delta>2G,\omega_{m}>\frac{\gamma_{m}}{2})$. The case that we
consider in this paper corresponds to
\begin{equation}\label{19}
\begin{array}{lcl}
\omega_{m}\gg\frac{\gamma_{m}}{2};\Delta>2G;\kappa\gg\gamma_{m};
\omega_{m}>\kappa.
\end{array}
\end{equation}
 The coupling between the normal modes would
be most efficient in the degenerate case i.e. when
$\omega_{m}=\sqrt{\Delta^2-4G^2}$. It is known from cavity QED that
the normal mode splitting leads to symmetric (asymmetric) spectra in
the degenerate (nondegenerate) case, provided that the dampings of
the individual modes are much smaller than the coupling constant.
Thus the mechanical oscillator is like the atomic oscillator, the
cavity mode in the rotating frame acquires the effective frequency
$\sqrt{\Delta^2-4G^2}$ which is dependent on the parametric
coupling. An estimate of the splitting can be made by using the
approximations given by Eq. (\ref{19}) and the zeroes of
$d(\omega)$. We find that the frequency splitting is given by
\footnote{This is derived by setting $\kappa$ and $\gamma_{m}$ in
$d(\omega)$ zero. A better estimate can be obtained by dropping
$i\gamma_{m}\omega$ and $\kappa^2$, but keeping the term
$-2i\kappa\omega$. This is because $\kappa/\omega_{m}(\simeq0.22)$
is not much smaller than 1.}
\begin{equation}\label{20}
\begin{array}{lcl}
\omega^{2}_{\pm}\cong\frac{\omega_{m}^2+\Delta^2-4G^2}{2}\pm\sqrt{(\frac{\omega_{m}^2-\Delta^2+4G^2}{2})^2+4\omega^2_{m}g^2},
\end{array}
\end{equation}
where we have defined
\begin{equation}\label{21}
\begin{array}{lcl}
g^{2}=\omega_{m}\chi^2|c_s|^2 [\Delta+2G\sin(\theta-2\varphi)],
\hspace{0.05in}e^{2i\varphi}=c_{s}^2/|c_{s}|^2.
\end{array}
\end{equation}
It should be borne in mind that $c_{s},\Delta$ etc are dependent on
the parametric coupling $G$. The splitting is determined by the pump
power, the couplings $\chi$ and $G$.

The parameters used are the same as those in the recent successful
experiment on optomechanical normal mode splitting
~\cite{Aspelmeyer}: the wave length of the laser $\lambda=2\pi
c/\omega_L=1064$ nm, $L=25$ mm, $m=145$ ng,
$\kappa=2\pi\times215\times10^3$ Hz,
$\omega_m=2\pi\times947\times10^3$ Hz, $ T=300$ mK, the mechanical
quality factor $Q^{'}=\omega_{m}/\gamma_{m}=6700$, parametric phase
$\theta=\pi/4$. And in the high temperature limit $k_B
T\gg\hbar\omega_{m}$, we have $\coth(\hbar\omega/2k_B T)\approx2k_B
T/\hbar\omega$.

\begin{figure}[htp]
 \scalebox{0.65}{\includegraphics{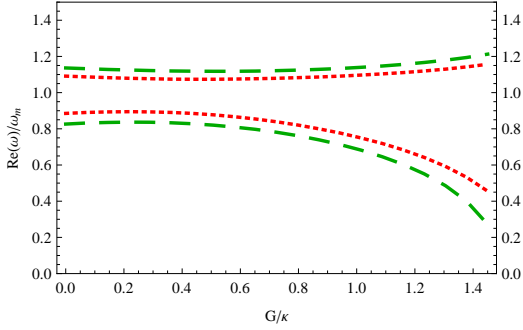}}
 \caption{\label{Fig2}(Color online)  The roots of $d(\omega)$ in the domain
Re$(\omega)>0$ as a function of parametric gain. $\wp=6.9$ mW
(dotted line), $\wp=10.7$ mW (dashed line). Parameters: the cavity
detuning $\Delta=\omega_{m}$.}
\end{figure}

\begin{figure}[htp]
 \scalebox{0.65}{\includegraphics{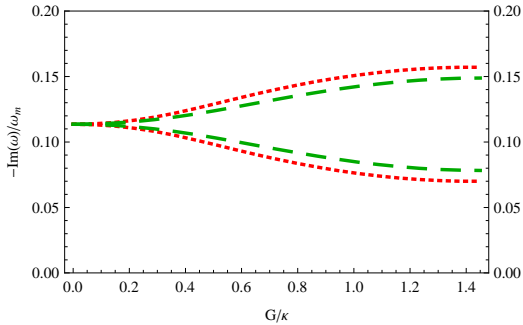}}
 \caption{\label{Fig3}(Color online)  The imaginary parts of the roots of $d(\omega)$
as a function of parametric gain. $\wp=6.9$ mW ( dotted line),
$\wp=10.7$ mW (dashed line). Parameters: the cavity detuning
$\Delta=\omega_{m}$.}
\end{figure}

Figure ~\ref{Fig2} shows the roots of $d(\omega)$ in the domain
Re$(\omega)>0$ for different values of $G$. Figure ~\ref{Fig3} shows
imaginary parts of the roots of $d(\omega)$ for different values of
$G$. The parametric coupling affects the width of the lines and this
for certain range of parameters aids in producing well split lines.
One root broadens and the other root narrows. The root that broadens
is the one that moves further away from the position for $G=0$.


\section{The spectra of the output field}
In this section, we would like to calculate the spectra of the
output field. The fluctuations $\delta c(\omega)$ of the cavity
field can be obtained from Eq. (\ref{8}). Further using the
input-output relation \footnote{See Ref. ~\cite{Walls}, p124, Eq.
(7.18).} $c_{out}(\omega)=\sqrt{2\kappa}c(\omega)-c_{in}(\omega)$,
the fluctuations of the output field are given by
\begin{equation}\label{22}
\begin{array}{lcl}
\delta c_{out}(\omega)=V(\omega)\xi(\omega)+E(\omega)\delta
c_{in}(\omega)+F(\omega)\delta c^{\dag}_{in}(-\omega),
\end{array}
\end{equation}
where
\begin{equation}\label{23}
\begin{array}{lcl}
V(\omega)=-\frac{\sqrt{2\kappa}\omega_{m}^2\chi}{d(\omega)}i\{[\kappa-i(\omega+\Delta)]c_{s}-2G
e^{i\theta}c_{s}^{*}\},\\
E(\omega)=\frac{2\kappa}{(\kappa-i\omega)^2+\Delta^2-4G^2}[-\frac{2\omega_{m}^3\chi^2}{d(\omega)}i\{[\kappa-i(\omega+\Delta)]c_{s}
\vspace{0.1in}\\\hspace{0.5in}-2Ge^{i\theta}c_{s}^{*}\}\{[\kappa-i(\omega+\Delta)]c_{s}^{*}+2Ge^{-i\theta}c_{s}\}\vspace{0.1in}\\\hspace{0.5in}+\kappa-i(\omega+\Delta)]-1,\\
F(\omega)=\frac{2\kappa}{(\kappa-i\omega)^2+\Delta^2-4G^2}[-\frac{2\omega_{m}^3\chi^2}{d(\omega)}i\{[\kappa-i(\omega+\Delta)]c_{s}
\vspace{0.1in}\\\hspace{0.5in}-2Ge^{i\theta}c_{s}^{*}\}\{[\kappa-i(\omega-\Delta)]c_{s}+2Ge^{i\theta}c_{s}^{*}\}\vspace{0.1in}\\\hspace{0.5in}+2Ge^{i\theta}].\\
\end{array}
\end{equation}
In Eq. (\ref{22}), the first term associated with $\xi(\omega)$
stems from the thermal noise of the mechanical oscillator, while the
other two terms are from the input vacuum noise. So the fluctuations
of the output field are influenced by the thermal noise and the
input vacuum noise.

The spectra of the output field are defined as
\begin{equation}\label{24}
\begin{array}{lcl}
\langle \delta c_{out}^{\dag}(-\Omega)\delta
c_{out}(\omega)\rangle=2\pi
S_{cout}(\omega)\delta(\omega+\Omega),\vspace{0.05in}\\
\langle \delta x_{out}(\Omega)\delta x_{out}(\omega)\rangle=2\pi
S_{xout}(\omega)\delta(\omega+\Omega),\vspace{0.05in}\\
\langle \delta y_{out}(\Omega)\delta y_{out}(\omega)\rangle=2\pi
S_{yout}(\omega)\delta(\omega+\Omega).
\end{array}
\end{equation}
where $\delta x_{out}(\omega)$ and $\delta y_{out}(\omega)$ are the
Fourier transform of the fluctuations $\delta x_{out}(t)$ and
$\delta y_{out}(t)$ of the output field , which are defined by
$\delta x_{out}(t)=\delta c_{out}(t)+\delta c_{out}^{\dag}(t)$ and
$\delta y_{out}(t)=i[\delta c_{out}^{\dag}(t)-\delta c_{out}(t)]$
~\cite{Walls}. Here $S_{cout}(\omega)$ denotes the spectral density
of the output field, $S_{xout}(\omega)$ means the spectrum of
fluctuations in the $x$ quadrature of the output field, and
$S_{yout}(\omega)$ is the spectrum of fluctuations in the $y$
quadrature of the output field.

Combining Eq. (\ref{16}), Eq. (\ref{22}), and
 Eq. (\ref{24}), we obtain the spectra of the output field
\begin{equation}\label{25}
\begin{array}{lcl}
S_{cout}(\omega)=V^{*}(\omega)V(\omega)\times2\frac{\gamma_{m}}{\omega_{m}}\omega[-1+\coth(\frac{\hbar\omega}{2k_{B}T})]\vspace{.1in}\\
\hspace{.7in}+F^{*}(\omega)F(\omega),\vspace{.1in}\\
S_{xout}(\omega)=[V(-\omega)+V^{*}(\omega)][V(\omega)+V^{*}(-\omega)]\vspace{.1in}\\
\hspace{.7in}\times2\frac{\gamma_{m}}{\omega_{m}}\omega[-1+\coth(\frac{\hbar\omega}{2k_{B}T})]\vspace{.1in}\\
\hspace{.7in}+[E(-\omega)+F^{*}(\omega)][F(\omega)+E^{*}(-\omega)],\vspace{.1in}\\
S_{yout}(\omega)=-[V^{*}(\omega)-V(-\omega)][V^{*}(-\omega)-V(\omega)]\vspace{.1in}\\
\hspace{.7in}\times2\frac{\gamma_{m}}{\omega_{m}}\omega[-1+\coth(\frac{\hbar\omega}{2k_{B}T})]\vspace{.1in}\\
\hspace{.7in}-[F^{*}(\omega)-E(-\omega)][E^{*}(-\omega)-F(\omega)].
\end{array}
\end{equation}
From Eq. (\ref{25}), it is seen that any spectrum of the output
field includes two terms, the first term is from the contribution of
the thermal noise of the mechanical oscillator, the second term is
from the contribution of the input vacuum noise.

We note that the spectra $S_{Q}(\omega)$, $S_{cout}(\omega)$,
$S_{xout}(\omega)$, and $S_{yout}(\omega)$ are determined by the
detuning $\Delta$, parametric gain $G$, parametric phase $\theta$,
input laser power $\wp$, and cavity length $L$. In the following we
will concentrate on discussing the dependence of the spectra on
parametric gain and input laser power.

\section{Numerical results}

In this section, we numerically evaluate the spectra
$S_{Q}(\omega)$, $S_{cout}(\omega)$, $S_{xout}(\omega)$, and
$S_{yout}(\omega)$ given by Eq. (\ref{17}) and Eq. (\ref{25}) to
show the effect of an OPA in the cavity on the normal mode splitting
of the movable mirror and the output field.

\begin{figure}[htp]
 \scalebox{0.65}{\includegraphics{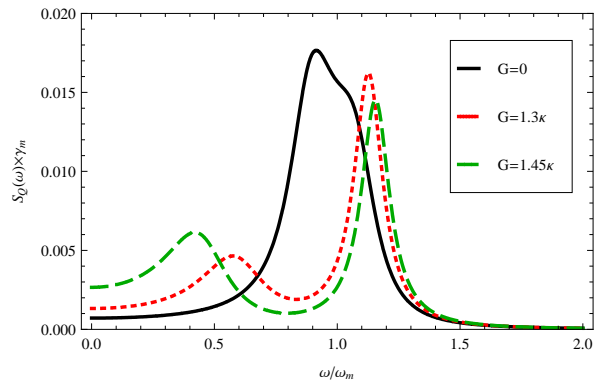}}
 \caption{\label{Fig4}(Color online)  The scaled spectrum $S_{Q}(\omega)\times \gamma_m$ versus
 the normalized frequency $\omega/\omega_{m}$ for different parametric gain. $G$= 0
(solid curve), 1.3$\kappa$ (dotted curve), 1.45$\kappa$ (dashed
curve). Parameters: the cavity detuning $\Delta=\omega_{m}$, the
laser power $\wp=6.9$ mW.}
 \end{figure}

\begin{figure}[htp]
 \scalebox{0.65}{\includegraphics{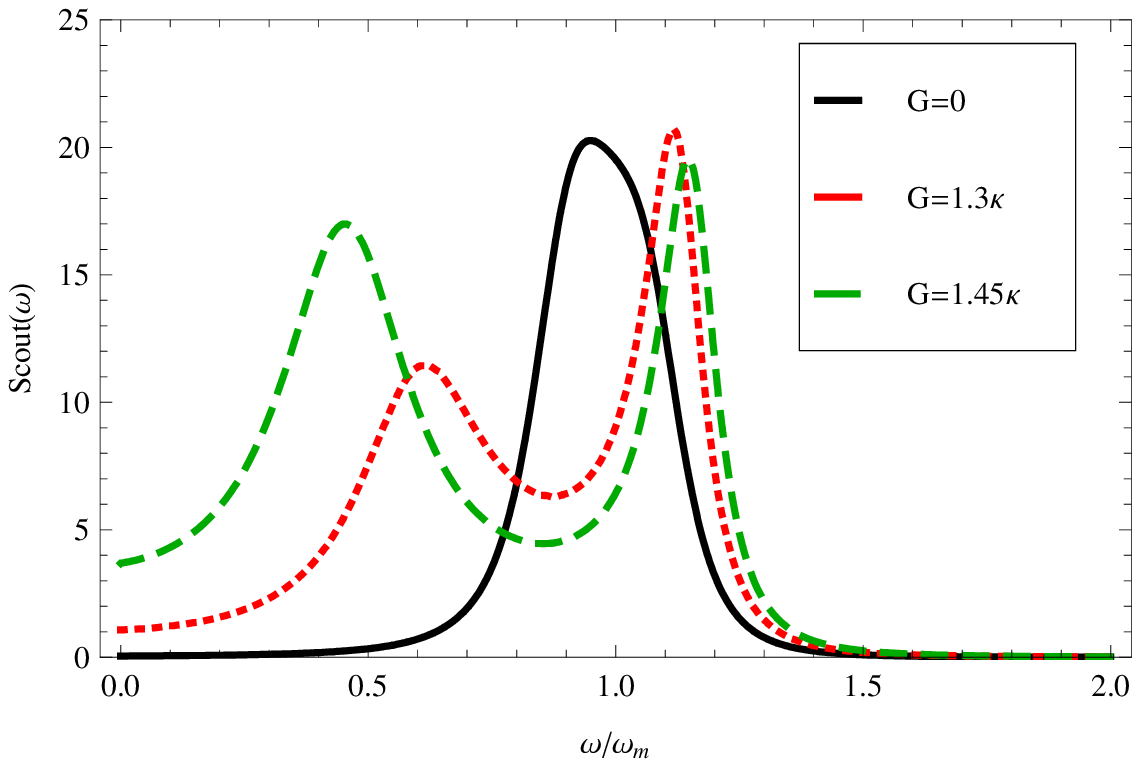}}
 \caption{\label{Fig5}(Color online)  The spectrum $S_{cout}(\omega)$ versus
 the normalized
frequency $\omega/\omega_{m}$ for different parametric gain. $G$= 0
(solid curve), 1.3$\kappa$ (dotted curve), 1.45$\kappa$ (dashed
curve). Parameters: the cavity detuning $\Delta=\omega_{m}$, the
laser power $\wp=6.9$ mW.}
 \end{figure}

\begin{figure}[htp]
 \scalebox{0.65}{\includegraphics{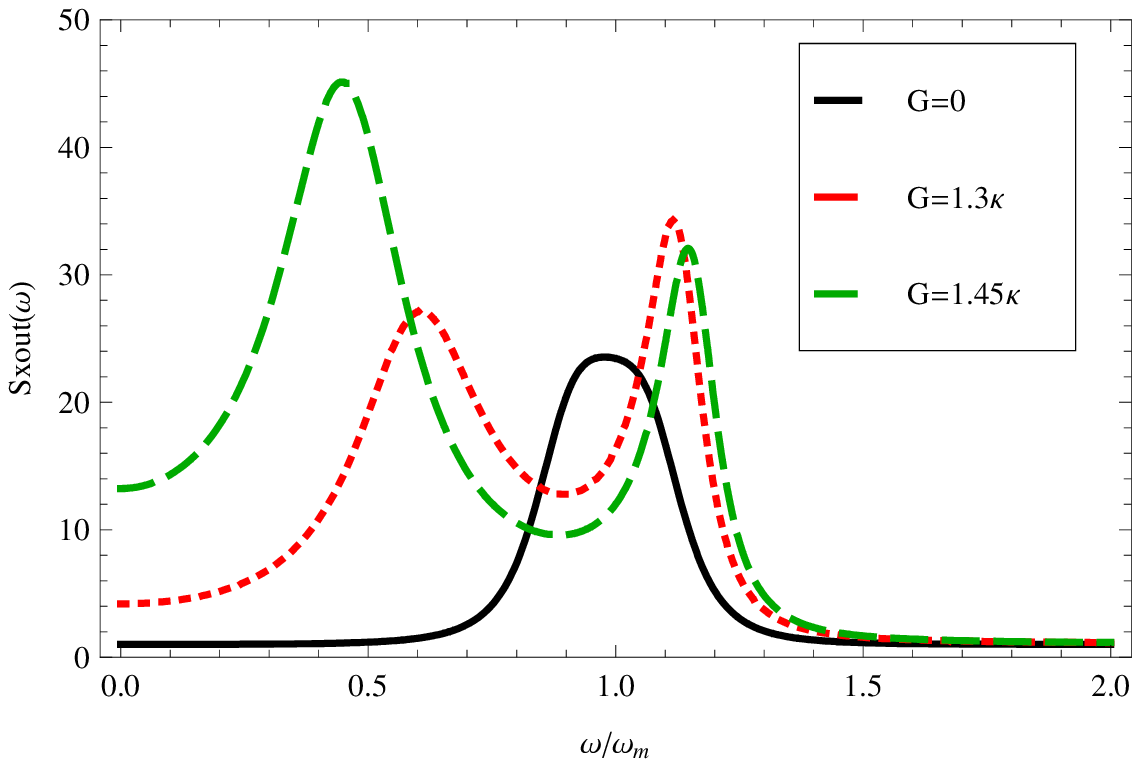}}
 \caption{\label{Fig6}(Color online)  The spectrum $S_{xout}(\omega)$ versus
 the normalized
frequency $\omega/\omega_{m}$ for different parametric gain. $G$= 0
(solid curve), 1.3$\kappa$ (dotted curve), 1.45$\kappa$ (dashed
curve). Parameters: the cavity detuning $\Delta=\omega_{m}$, the
laser power $\wp=6.9$ mW.}
\end{figure}

\begin{figure}[htp]
 \scalebox{0.65}{\includegraphics{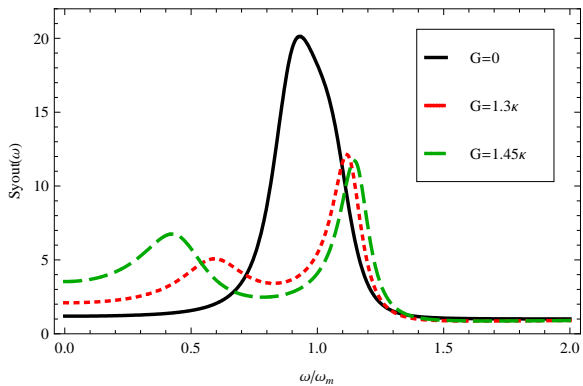}}
 \caption{\label{Fig7}(Color online) The spectrum $S_{yout}(\omega)$ versus
 the normalized
frequency $\omega/\omega_{m}$ for different parametric gain. $G$= 0
(solid curve), 1.3$\kappa$ (dotted curve), 1.45$\kappa$ (dashed
curve). Parameters: the cavity detuning $\Delta=\omega_{m}$, the
laser power $\wp=6.9$ mW.}
 \end{figure}

We consider the degenerate case $\Delta=\omega_{m}$ for $G=0$, and
choose $\wp=6.9$ mW to satisfy the stability conditions (12),
parametric gain must satisfy $G\le1.62\kappa$. Figures ~\ref{Fig4}
--~\ref{Fig7} shows the spectra $S_{Q}(\omega)$, $S_{cout}(\omega)$,
$S_{xout}(\omega)$, and $S_{yout}(\omega)$ as a function of the
normalized frequency $\omega/\omega_{m}$ for various values of
parametric gain. When the OPA is absent ($G=0$), the spectra barely
show the normal mode splitting. As parametric gain is increased, the
normal mode splitting becomes observable. This is due to significant
changes in the line widths and position. When $G=1.3\kappa$, two
peaks can be found in the spectra. Note that the separation between
two peaks becomes larger as parametric gain increases. The reason is
that increasing the parametric gain causes a stronger coupling
between the movable mirror and the cavity field due to an increase
in the photon number in the cavity. The values of intercavity photon
number $|c_{s}|^{2}$ are $2.68\times10^{9}$, $4.30\times10^{9}$,
$5.65\times10^{9}$ for $G=0$, $1.3\kappa$, and $1.45\kappa$
respectively. We have examined the contributions of various terms in
Eq. (\ref{25}) to the output spectrum. The dominant contribution
comes from the mechanical oscillator. Note further the similarity
~\cite{Aspelmeyer} of the spectrum of the output quadrature $y$
(Fig. ~\ref{Fig7}) to the spectrum of the mechanical oscillator
(Fig. ~\ref{Fig4}). It should be borne in mind that the strong
asymmetries in the spectra for $G\neq0$ arise from the fact that by
fixing $\Delta$ at $\omega_{m}$, the frequencies of the cavity mode
and the mechanical oscillator do not coincide if $G\neq0$; $\chi=0$.
Besides the damping term $\kappa$, $\kappa$ being not negligible
compared to $\Delta$, also contributes to asymmetries.

\begin{figure}[htp]
 \scalebox{0.65}{\includegraphics{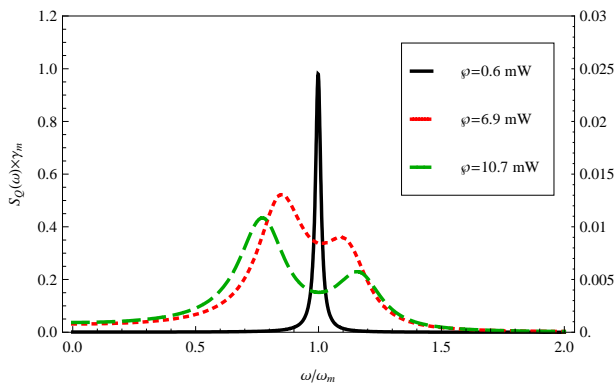}}
 \caption{\label{Fig8}(Color online)  The scaled spectrum $S_{Q}(\omega)\times\gamma_m$ versus
 the normalized
frequency $\omega/\omega_{m}$, each curve corresponds to a different
input laser power. $\wp$= 0.6 mW (solid curve, leftmost vertical
scale), 6.9 mW (dotted curve, rightmost vertical scale), 10.7 mW
(dashed curve, rightmost vertical scale). Parameters: the cavity
detuning $\Delta=\sqrt{\omega_{m}^2+4G^2}$, parametric gain
$G=1.3\kappa$.}
 \end{figure}

 Now we fix
parametric gain $ G=1.3\kappa $, and choose
$\Delta=\sqrt{\omega_{m}^2+4G^2}$, the input laser power must
satisfy $\wp\le55$ mW.  The spectrum $S_{Q}(\omega)$ as a function
of the normalized frequency $\omega/\omega_{m}$ for increasing the
input laser power is shown in Fig.~\ref{Fig8}. As we increase the
laser power from 0.6 mW to 10.7 mW, the spectrum exhibits a doublet
and the peak separation is proportional to the laser power, because
the coupling between the movable mirror and the cavity field for a
given parametric gain $G$ is increased with increasing the input
laser power due to an increase in photon number.

\begin{figure}[htp]
 \scalebox{0.65}{\includegraphics{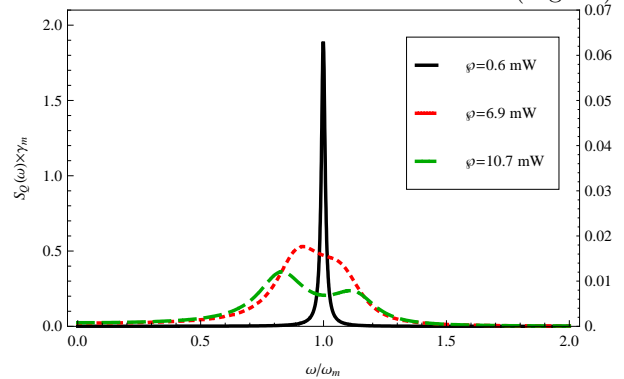}}
 \caption{\label{Fig9}(Color online)  The scaled spectrum $S_{Q}(\omega)\times \gamma_m$ versus
 the normalized
frequency $\omega/\omega_{m}$, each curve corresponds to a different
input laser power. $\wp$= 0.6 mW (solid curve, leftmost vertical
scale), 6.9 mW (dotted curve, rightmost vertical scale), 10.7 mW
(dashed curve, rightmost vertical scale). Parameters: the cavity
detuning $\Delta=\omega_{m}$, parametric gain $G=0$.}
 \end{figure}

For comparison, we also consider the case of the cavity without OPA
($G=0 $), the spectrum $S_{Q}(\omega)$ as a function of the
normalized frequency $\omega/\omega_{m}$ for increasing the input
laser power at $\Delta=\omega_{m}$ is plotted in Fig. ~\ref{Fig9}.
We can see if the laser power is increased from 0.6 mW to 10.7 mW,
the spectrum also displays normal mode splitting. However the normal
mode with OPA (Fig. ~\ref{Fig8}) are more pronounced than that in
the absence of OPA (Fig. ~\ref{Fig9}).

\section{Conclusions}
In conclusion, we have shown how the normal mode splitting behavior
of the movable mirror and the output field is affected by the OPA in
the cavity. We work in the resolved sideband regime and operate
under the stability conditions (12). We find that increasing
parametric gain can make the spectra $S_{Q}(\omega)$,
$S_{cout}(\omega)$, $S_{xout}(\omega)$, and $S_{yout}(\omega)$
evolve from a single peak to two peaks. Furthermore, for a given
parametric gain, increasing input laser power can increase  the
amount of normal mode splitting of the movable mirror due to the
stronger coupling between the movable mirror and the cavity field.

We thank Dr. M. Aspelmeyer for bringing the experimental paper
~\cite{Aspelmeyer} to our notice and for interesting correspondence.
We gratefully acknowledge support from the NSF Grant No. PHYS
0653494.

\end{document}